# Machine learning and structure formation in modified gravity

Jonathan C. Betts,[1]★ Carsten van de Bruck,[1] Christian Arnold [2] and Baojiu Li [2]

[1]*School of Mathematics and Statistics, University of Sheffield, Hounsfield Road, S3 7RH Sheffield, United Kingdom*
[2]*Institute for Computational Cosmology, Department of Physics, Durham University, South Road, Durham DH1 3LE, United Kingdom*

## ABSTRACT
In general relativity, approximations based on the spherical collapse model such as Press–Schechter theory and its extensions are able to predict the number of objects of a certain mass in a given volume. In this paper, we use a machine learning algorithm to test whether such approximations hold in screened modified gravity theories. To this end, we train random forest classifiers on data from *N*-body simulations to study the formation of structures in lambda cold dark matter (ΛCDM) as well as screened modified gravity theories, in particular $f(R)$ and nDGP gravity. The models are taught to distinguish structure membership in the final conditions from spherical aggregations of density field behaviour in the initial conditions. We examine the differences between machine learning models that have learned structure formation from each gravity, as well as the model that has learned from ΛCDM. We also test the generalizability of the ΛCDM model on data from $f(R)$ and nDGP gravities of varying strengths, and therefore the generalizability of extended Press–Schechter spherical collapse to these types of modified gravity.

**Key words:** dark energy – dark matter – large-scale structure of Universe.

## 1 INTRODUCTION

Structure formation in the Universe is caused by gravitational instability due to tiny density fluctuations produced in the very early Universe. It proceeds from the bottom up, with objects smaller than galaxies forming first, which subsequently form larger structures such as galaxies and clusters of galaxies. An important ingredient in this theory is the existence of dark matter (DM), which allows structures to form on small scales despite dissipative processes in baryonic matter. An essential element in this picture of structure formation are DM haloes, which can be thought of as the building blocks of the structures in the Universe. These are gravitationally bound objects and understanding their formation and evolution is an essential goal in cosmology. Due to the fact that DM haloes are highly non-linear objects, one has to resort to *N*-body simulations to study the full evolution of DM haloes. Despite this, there are (semi-)analytic approximations, which allow to estimate the redshift at which structures go non-linear or to calculate the halo mass function. The first study was due to Press & Schechter (1974), with subsequent improvements made in e.g. Sheth & Tormen (1999), Jenkins et al. (2001), and Tinker et al. (2008).

These investigations rely on general relativity (GR), in the sense that gravity alone is responsible for structure formation. However, ever since the discovery of the accelerated expansion of the Universe (Riess, Filippenko et al. 1998; Perlmutter et al. 1999), cosmologists have studied various theories accounting for the late time accelerated expansion. In the simplest models, a new energy form, dubbed dark energy (DE), is responsible for the accelerated expansion. Including a cosmological constant in GR is the most economical way to model DE. Adding the cosmological constant in GR, together with DM, leads to a highly successful model [the lambda cold dark matter (ΛCDM) model], which is an agreement with many of the current data (Peebles & Ratra 2003; Abbott et al. 2019; Peebles 2020). However, the cosmological constant remains an enigma. Its value and physical interpretation is currently unexplained. Models of DE such as quintessence or theories of modified gravity have been developed in order to explain current data and to challenge the ΛCDM model. In models of modified gravity in particular, new degrees of freedom couple to matter, resulting in an additional force affecting structure formation in the Universe. In such theories, the analytical models such as the Press–Schechter model mentioned above, might be in need of modifications (Pace, Waizmann & Bartelmann 2010; von Braun-Bates & Devriendt 2018).

In this paper, we will use a machine learning approach to gain insight into structure formation in modified gravity theories. We use the approach put forward in Lucie-Smith et al. (2018), in which the evolution of structure formation was turned into a (binary) classification problem. In this approach, a random forest classifier is trained to learn the relationship between the initial conditions of the DM density field (at high-redshift) and the final DM distribution. The initial conditions are mapped on to whether a particle ends up in a halo or not. The random-forest algorithm has been shown to be very effective and accurate in Lucie-Smith et al. (2018). As it was shown in that paper, the predictions of the random-forest classifier are in very good agreement with extended Press–Schechter theory. The question we are addressing in this paper is whether this mapping from initial to final conditions is affected in theories of modified gravity. To be concrete, we are using data from simulations of $f(R)$ theories and nDGP (Dvali–Gabadaze–Porrati) gravity and perform the analysis of Lucie-Smith et al. (2018) on these data. If our result would show that the predictions get significantly worse as we move away from GR, then information about the additional degree of freedoms in modified gravity theories would need to be taken into account.

★ E-mail: jbetts3@sheffield.ac.uk

The paper is organized as follows. In Section 2, we will briefly summarize the modified gravity theories we consider in this paper. In Section 3, we summarize the simulations. Our methodology is presented in Section 4 and our results are presented in Section 5. Section 6 summarizes our findings.

## 2 MODIFIED GRAVITY THEORIES

In this paper, we consider a couple of modified gravity theories, namely a $f(R)$ model and the nDGP model. We will briefly describe these two models to set the scene.

### 2.1 $f(R)$ gravity

In this extension of GR, an additional term is added to the Einstein–Hilbert action. The theory is described by the following action

$$S = \int d^4x \sqrt{-g} \left[ \frac{R + f(R)}{16\pi G_N} + \mathcal{L}_M \right]. \quad (1)$$

In this equation, $G_N$ is Newton's gravitational constant, $\mathcal{L}_M$ is the matter Lagrangian (including the standard model fields and CDM). The function $f(R)$ encodes the modifications from GR. In this paper, we will concentrate on a specific form of $f(R)$, proposed by Hu and Sawicki (Hu & Sawicki 2007). This takes the form

$$f(R) = -m^2 \frac{c_1 \left(-R/m^2\right)^n}{c_2 \left(-R/m^2\right)^n + 1}, \quad (2)$$

where $m^2$ sets the DE energy scale and is given by $m^2 \equiv H_0^2 \Omega_M$. The parameter $c_1$, $c_2$, and $n$ are free parameters, which are chosen such that the theory matches the $\Lambda$CDM background evolution. Apart from the metric, $f(R)$ theories contain an additional scalar degree of freedom, the scalaron $f_R \equiv df/dR$. This degree of freedom is coupled universally to all matter forms and mediates a new force. To make $f(R)$ theories compatible with experiments, one can employ the chameleon mechanism, for which it is necessary that $|f_R| \ll 1$ in high curvature regions. The scalaron in this model always sits at a minimum of an effective potential, which leads to a relationship between the parameter and the value of the scalaron today (Arnold, Leo & Li 2019b)

$$\frac{c_1}{c_2^2} = -\frac{1}{n} \left[ 3 \left( 1 + 4\frac{\Omega_\Lambda}{\Omega_M} \right) \right] f_{R0}. \quad (3)$$

The original Hu–Sawicki formulation uses theoretical arguments for screening within our Solar system to place a constraint of $|f_{R0}| \lesssim 10^{-6}$. Other constraints from distance indicators in the nearby Universe (Jain, Vikram & Sakstein 2012), dwarf galaxies (Vikram et al. 2013) lower this value even further. However, the weakness of modifications that meet these constraints would make their impact on structure formation on cosmological scales negligible as they are not permitted to impact clustering of large-scale structure (Liu et al. 2021). Despite these shortcomings, effective models in which the fifth force can impact spherical collapse still provide a test bed for gravity on large scales. More recently, clustering constraints on massive haloes (Cataneo et al. 2015) place $|f_{R0}| < 10^{-4.79}$ in these effective models.

We work with $N$-body simulation data for which $n = 1$ and $|f_{R0}| = 10^{-4}, 10^{-5}, 10^{-6}$ were chosen. In the following, these models are denoted by F4, F5, and F6, respectively.

### 2.2 nDGP gravity

The DGP model is motivated from theories with extra dimensions (Dvali, Gabadadze & Porrati 2000). In this model, our Universe is a brane embedded in a five-dimensional space–time, called the bulk. The action is given by

$$S = \int_{\text{bulk}} d^5x \sqrt{-g^{(5)}} \frac{R^{(5)}}{16\pi G^{(5)}} + \int_{\text{brane}} d^4x \sqrt{-g} \frac{R}{16\pi G}. \quad (4)$$

In this equation, $g^{(5)}$ denotes the determinant of the metric of the five-dimensional space–time, $g$ is the determinant of the induced metric on the brane. $G$ is Newton's gravitational constant and $G^{(5)}$ is the five-dimensional gravitational coupling. Since there are two gravitational couplings in the theory, their ratio define a characteristic scale, called cross-over scale, defined by the ratio

$$r_c = \frac{G^{(5)}}{2G},$$

above which the first term in the action above dominates and deviations from GR are predicted. The model consists of two branches. The first branch, considered here, is the 'normal' branch (nDGP), and the second branch is a self-accelerating branch. We focus on the former, because ghost instabilities are not present in this branch. Above the cross-overscale, gravity becomes stronger, while on smaller scales gravity behaves like GR thanks to Vainshstein-screening.

For the models used in this work, we express deviations from GR by the product $H_0 r_c$ (i.e. the ratio of $r_c$ to the present Hubble radius $1/H_0$). Markov chain Monte Carlo studies of brane tension (Lombriser et al. 2009) place the crossover scale much higher than the Hubble scale, giving the constraint $H_0 r_c > 3$. For the models considered in this paper, we use $H_0 r_c = 1$ and $H_0 r_c = 5$, denoted by N1 and N5, respectively.

## 3 SIMULATIONS

In this work, we make use of the SHYBONE (Simulation HYdrodynamics BeyONd Einstein) simulation suite (Arnold, Leo & Li 2019a; Hernández-Aguayo et al. 2021), which are a set of high-resolution hydrodynamical galaxy formation simulations of GR, $f(R)$ gravity, and the nDGP model, carried out with the AREPO hydrodynamical simulation code (Springel 2010), modified to include a modified gravity solver that solves the extra scalar fields in the $f(R)$ and nDGP models. For this analysis only the DM runs (with no baryons) of SHYBONE are used.

These simulations cover six gravity models – GR, F6, F5, F4, N5, and N1 – and evolve $512^3$ DM particles in a cubic box of size $62\ h^{-1}$ Mpc from $z = 129$ to $z = 0$. All runs start from identical initial conditions, given that the modified gravity impact on matter clustering at $z = 129$ is negligible. Group catalogues are identified using the SUBFIND (Springel et al. 2001) halo-finder, and we use $M_{200c}$ as the halo mass definition, which includes all particles enclosed by a sphere of radius $R_{200c}$, within which the average density is $\bar{\rho} = 200 \times \rho_{\text{crit}}$ around the potential minimum of the object, at the halo redshift.

The simulations adopt the Planck 2015 cosmology (Planck Collaboration 2016) with cosmological parameters $n_s = 0.9667$, $h = 0.6774$, $\Omega_\Lambda = 0.6911$, $\Omega_b = 0.0486$, $\Omega_m = 0.3089$, and $\sigma_8 = 0.8159$, where $h \equiv H_0/\left(100\ \text{km s}^{-1}\ \text{Mpc}^{-1}\right)$, $n_s$ and $\sigma_8$ are respectively the spectral index of the primordial density fluctuations and the present-day root-mean-squared matter density fluctuation within $8\ h^{-1}$ Mpc, and $\Omega_b$, $\Omega_m$, and $\Omega_\Lambda$ are the present-day density parameters of baryons, (all) matter and the cosmological constant. The particle mass resolution is $\simeq 1.5 \times 10^8 h^{-1}$ M$_\odot$.

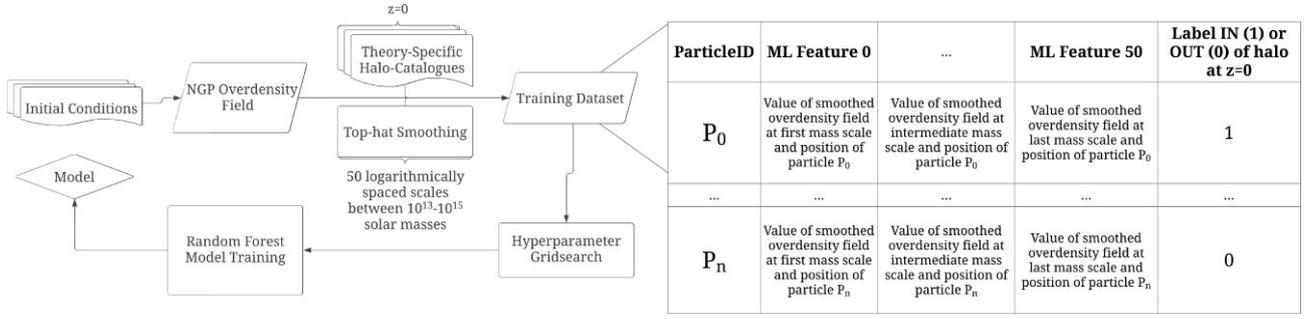

**Figure 1.** The full pipeline used to generate models for GR, three *f(R)* theories, and two nDGP theories. The initial conditions of each simulation are used to generate an NGP density contrast field image. This density image then has the feature engineering smoothing applied to generate 50 features per particle describing its density environment. The halo catalogues at *z* = 0 are then used to label the features for supervised training. Finally, the resulting data sets are used to train a random forest classifier per theory, by implementing a grid search to optimize hyperparameters.

## 4 MACHINE LEARNING METHODS

The following design of a machine learning pipeline to study structure membership and its influences in modified gravity is proposed; following the methods of Lucie-Smith et al. (2018). The *N*-body simulations of Section 3 are used to generate supervised training data sets for models that can classify the structure membership of particles at $z = 0$ based on initial density behaviour. The properties of these models can then be examined to study the generalizability of density-contrast-centric descriptions of structure formation (Press & Schechter 1974) to *f(R)* and nDGP screened gravity theories. Fig. 1 shows a high-level overview of this pipeline.

Supervised training data sets consist of a set of objects with a class membership label and a set of *features* for each object that are designed to contain relevant information to class membership. In this case, the objects are particles, they are given IN/OUT labels based on structure membership at $z = 0$ and the features consist of the value of the density contrast centred on the position of that particle in the initial conditions smoothed at increasingly non-local mass scales. Models trained on such data can then be examined to determine which of the features was most relevant to classification output, and therefore how much structure membership information is contained in the density contrast and at which scale.[1]

### 4.1 Structure membership

There are several available halo-finder algorithms that group the masses of an *N*-body simulation into a catalogue of haloes at a given redshift slice. In order to define the IN and OUT classes, the SUBFIND halo catalogue at $z = 0$ is used as described in Section 3. Whilst these types of halo-finders are able to identify substructures in DM haloes, the binary nature of the problem means that only halo membership, rather than substructure membership is relevant in this case.

Particles are classified as being IN a structure if they are gravitationally bound to a halo of mass $M_{\text{halo}} \geq 1.8 \times 10^{12}$ M$_\odot$ and OUT if they are in no structure at all or in a halo of mass lower than this threshold value. This value is chosen at a typical scale in order to divide the halo catalogue into two binary classes, and may depend on the resolution of the *N*-body simulation being considered.[2] More thresholds could also be considered for a multiclass rather than a binary structure membership problem. The additional condition of gravitational binding preserves the assumptions of spherical collapse into haloes and places particles in extraneous filaments in the OUT class.

We emphasize at this point that gravitational binding in our halofinder is defined as in GR (using the comparison of the kinetic energy of a given particle to the Newtonion potential of the associated structure) irrespective of the modified gravity model. We find that the impact of this assumption is negligible on our results, a more detailed discussion can be found in Appendix A.

### 4.2 Feature engineering

Now the local density behaviour around the particle in the initial conditions must be turned into features that differentiate between particles that end up IN or OUT of haloes. The base density contrast field is generated using the nearest-grid-point (NGP) method.[3] This field is then spherically smoothed to generate features as follows.

For a given smoothing radius *R*, the smoothed density contrast can be defined as

$$\delta(\mathbf{x}; R) = \int \delta(\mathbf{x}') W_{\text{top-hat}}(\mathbf{x} - \mathbf{x}'; R) d^3 \mathbf{x}', \quad (5)$$

with real-space top-hat window function

$$W_{\text{top-hat}} = \begin{cases} \frac{3}{4\pi R^3} & \text{for } |\mathbf{x}| \leq R, \\ 0 & \text{for } |\mathbf{x}| > R. \end{cases} \quad (6)$$

Convolution (5) is calculated for a given *R* by means of a Fourier transform, filtering in *k* space, then inverse Fourier transform. $W_{\text{top-hat}}(\mathbf{x}, R)$ for a given *R* maps to a characteristic mass scale $M_{\text{smoothing}} = \bar{\rho} V_{\text{top-hat}}(R)$ with $V_{\text{top-hat}}(R) = (4//3)\pi R^3$. Each feature for a given particle then becomes the smoothed density contrast value for a chosen $M_{\text{smoothing}}$ centred on the particle in the initial conditions. The full feature set is computed for a series of logarithmically spaced values of $M_{\rm smoothing}$ bounded below by the resolution of the simulation and bounded above by the size of the simulation box. In the case of this work, $3 \times 10^{10}\,{\rm M}_\odot < M_{\rm smoothing} < 10^{15}\,{\rm M}_\odot$. It has been found previously that a series of 50 values for $M_{\rm smoothing}$ in this range and therefore, 50 features was sufficient to differentiate class membership (Lucie-Smith et al. 2018). The same is true in this work, a larger number of features did not yield an improvement in classification performance for GR or any of F4, F5, F6, N1, and N5. The set of 50 smoothed density contrast features for a given particle can be visualized as its density trajectory through increasing mass scales. Fig. 2 shows a selection of example particle trajectories for the IN and OUT classes in GR, and Fig. 3 shows the averaged trajectories for each class. This demonstrates how the features given to the classifier can be used to differentiate class membership.

---

[1]This work makes use of the PYTHON programming framework. The PYNBODY library (Pontzen et al. 2013) and the AREPO PYTHON integration (Arnold et al. 2019a) are used to parse the simulation data. The PYLIANS CYTHON library is used to carry out the computation of the base density contrast field and to carry out smoothing convolution 5. The SK-LEARN machine learning library for PYTHON is used to carry out model training, inference, and analysis.

[2]Halo-mass-function cutoffs due to unresolved haloes at smaller scales are a good indicator for the structure threshold.

[3]Cloud-in-cell methods were also tested in the generation of the base density contrast field but were found to produce resolution effects when smoothed at low scales due to the boundaries between cells.

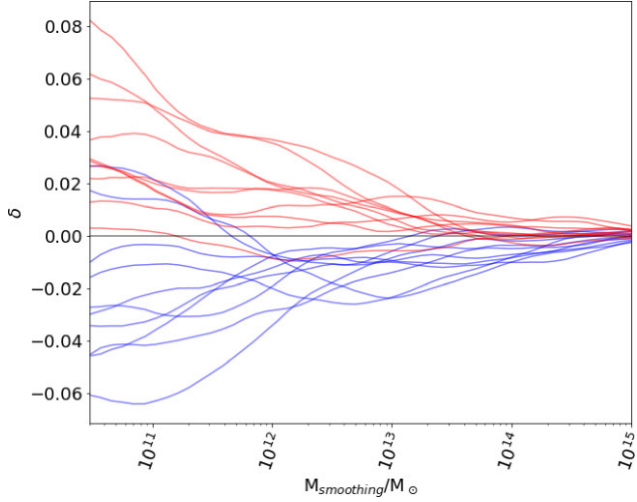

**Figure 2.** Density trajectories for example particles contained in the IN (red) and OUT (blue) classes for GR.

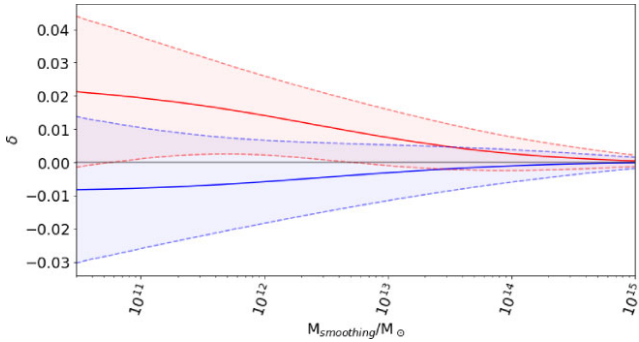

**Figure 3.** The mean trajectories for IN (red) and OUT (blue) in GR over a training data sample of 50 000 particles. The shaded area shows the 1$\sigma$ deviation in each class.

### 4.3 Model selection and training

There are many machine learning (ML) algorithms that can learn classification. For structure formation, each of the density contrast features are correlated with one another. Random-forest classifiers were found to be agnostic to this (Lucie-Smith et al. 2018), and we continue their use in this work for their auditability. Random forests are an ensemble model that use the wisdom of crowds principle to vote on the outcome of a classification amongst a forest of decision trees (Breiman et al. 1984; Breiman 2001). Decision tree methods attempt to divide up the feature space such that members of each class can be identified based on their feature values. Building the perfect single decision tree is computationally intractable in non-trivial cases (Géron 2019) and the resulting model would not generalize well to new data. Therefore a 'forest' of trees trained on random subsets of the training data vote on classification outcomes. This reduces bias and variance in predictions and increases the generalizability of the model.

Secondary among the benefits of using a random forest is the ability to examine the *feature importance* attributes of a model. Once the model has been trained, it is possible to measure each feature's importance to the classification decision. This is done by computing an average class-impurity decrease across nodes in the random forest that use the feature. We make use of *Gini* impurity (Farris 2010) in this work. Consider node $m$ of a given decision tree in the forest. Let the data at node $m$ be denoted by $Q_{\rm m}$ and contain $n_{\rm m}$ samples. The goal of node $m$ is to split each object $y \in Q_{\rm m}$ into one of $k$ classes. In this binary case $k \in [0, 1]$. The proportion of class $k$ observations at node $m$ is

$$p_{\rm mk} = \frac{1}{n_{\rm m}} \sum_{y \in Q_{\rm m}} I(y = k), \tag{7}$$

which is then aggregated into the Gini impurity measure for the node,

$$H_{\rm gini}(Q_{\rm m}) = \sum_k p_{\rm mk}(1 - p_{\rm mk}). \tag{8}$$

For feature $i$ in a single decision tree, $N_{\rm i}$ is the set of nodes that use feature $i$ and $N_{\rm t}$ is the total number of nodes in the tree. Feature importance for feature $i$ is then given by

$$F_{\rm i} \equiv \frac{\sum_{n \in N_{\rm i}} H_{\rm gini}(Q_{\rm n})}{\sum_{m \in N_{\rm t}} H_{\rm gini}(Q_{\rm m})}. \tag{9}$$

In the random forest case, $F_{\rm i}$ is computed for each feature, for each tree in the forest. These values are then averaged across all of the trees to give each feature an important statistic for the forest as a whole. Finally, the forest feature importance are normalized between 0 and 1. In this work, where each feature denotes the behaviour of the density contrast at a given mass scale centred on a particle, the feature importance should be thought of as the significance of density field behaviour at that mass scale to the process of structure formation.

Aside from the calculation of relevant features to describe objects to be classified, there are a couple of other nuances to training a machine learning model of this type that are worth mentioning. Parameters that give the algorithm computational information such as the number of trees in a forest, the maximum depth of each tree or the minimum number of training samples a node must take before it can finalize its split of the data set. These values are known as *hyperparameters*. The selection of these values can have a considerable effect on the performance of the resulting model, e.g. *overfitting*. Overfitting occurs when a model is too strongly biased towards the training data it has received and therefore performs poorly when asked to classify new instances when validated. Voting architectures such as the one used in this work can contribute somewhat to combatting this issue, as the biases of individual trees are smoothed out (Breiman 1996). However, there are also *cross-validation* (Stone 1974) techniques that can be used during training to help select hyperparameters that enable a model to learn and generalize well. The first method used in this work is a manipulation of the training data structure known as *k-folding* (Mosteller & Tukey 1968). k-Folding involves rotating the training data set through a train/validate process. Training data is divided into *k* subsets of equal size, the model is trained on *k* − 1 of the subsets and then validated on the leftover one. This process is then repeated with a different *k* − 1 subsets and so forth until each subset has been used to validate. The second method involves a higher level iteration over sets of algorithm hyperparameters known as a grid-search. A grid-search defines a hyperparameter space, samples from it, trains a model with the selected hyperparameters and then tests its performance against a validation data set. This carries on iteratively and the validation scores of all the models are compared to select the one that performed the best and therefore uses the best set of hyperparameters. The sampling of the hyperparameter space can either be done exhaustively or by random sampling, and can also be repeated itself if trends in performance emerge.

In this work, we will focus on four measures of model performance: the area under the receiver-operating-curve, precision, recall, and the F1 score (Bradley 1997; Olson & Delen 2008). The receiver-operating-curve (sometimes known as the receiver-operating-characteristic, abbreviated to ROC or ROC curve) is the primary evaluation method for classification problems. It is an aggregate of the *confusion matrix* output by a given model on a test set of data with known class labels. The confusion matrix groups the predictions of the model into *true positives*, *false positives*, *true negatives*, and *false negatives* for a given decision threshold. For example, take our two example classes $y_0$ and $y_1$. In classification matrix terms for a binary decision, let $y_0$ be considered the *negative* class and $y_1$ the *positive* class. For feature vector **x** the model will output class probabilities P($y_0$) and P($y_1$). If the decision threshold is set at 0.5, then P($y_1$) > 0.5 will result in a classification of the object with properties **x** into class $y_1$. If this prediction is true then the result is a true positive at decision threshold 0.5. If false, the prediction is a false positive. Vice versa for the negative class $y_0$. To generate the ROC, the fraction of positive predictions that are true (true positive rate) and the fraction of false positives to negatives (false positive rate) are computed at a series of decision thresholds between 0 and 1. The true positive rate is then plotted against the false positive rate, this is the ROC. The area under the resulting curve (AUC) forms a useful evaluation metric for models on a scale between 0 and 1, with 1 being a perfect score.

There are some limitations to this measure, such as its lack of robustness to class imbalance. If the training sample is 90 per cent $y_1$ objects, then classifying all objects as $y_1$ spikes the true positive rate, even if the performance on the negative class is poor. This leads to a misleadingly high AUC score. Fortunately, the other measures listed above in combination with the AUC score give a more complete picture of a classifier's performance. Precision and recall are defined individually, but are often used together to provide class-relative insight into a model. They are both also derived from the confusion matrix at a given decision threshold. Precision *P* is defined as

$$P \equiv \frac{T_p}{T_p + F_p}, \quad (10)$$

where $T_p$ is the number of true positives and $F_p$ is the number of false positives. Recall *R* is defined as

$$R \equiv \frac{T_p}{T_p + F_n}, \quad (11)$$

where $F_n$ is the number of false negatives. Semantically, precision describes the fraction of positive classifications that are true and recall gives the fraction of ground truth positive results that are correctly identified by the model. This is a subtle but important distinction. A precise model with low recall may not identify all of the instances of a class, but those members of a class it does identify are likely to be correctly classified. Conversely, a high recall but low precision model will catch most of the instances of a class, but will be less certain about the correctness of those predictions. Either case may be desirable depending on the problem. The final performance measure used in this work, the $F_1$ score, is defined as the harmonic mean of precision and recall

$$F_1 \equiv \frac{2PR}{P + R}. \quad (12)$$

$F_1$ scores also scale between 0 and 1, and are often used as part of a broader assessment of models that are attempting to solve class-imbalanced problems. Given the problem addressed in this work involves classifying particles inside and outside *N*-body scale structures, it follows that the class of particles that are in structures will be smaller than the class of particles outside of structures. A combination of AUC and $F_1$ scores will be used for grid-search assessment of hyperparameters.

A series of training data sets each containing 50 000 particles[4] is generated for each of GR, F4, F5, F6, N1, and N5. The first will be used for the k-folded gridsearch for hyperparameters described above. The rest of the data sets will be used to train a series of models using the selected hyperparameters for the given gravitational mechanism. The feature importance of these models will be collated to show how well the density contrast determines structure membership for a particle, and therefore how well it should govern analytical structure formation for the relevant modified gravity.

## 5 RESULTS

### 5.1 Individual model properties

Fig. 4 shows the feature importance attributes of models constructed for each cosmology and their respective mean AUC score performance. For our GR simulation not only does the density contrast provide a complete description of structure formation – as is expected – but most of the information is concentrated at $5.9 \times 10^{11}$ M$_\odot$. As the strength of the *f(R)* modifications increases away from GR (i.e. structures become less screened from modifications), the most relevant scale for structure classification moves inwards to $3.8 \times 10^{11}$ M$_\odot$. Similarly for nDGP gravity, as the enhancement of gravity gets stronger, the information also moves inwards from the range $3.8 \times 10^{11}$ M$_\odot$ < $M_{\text{smoothing}}$/M$_\odot$ < $5.9 \times 10^{11}$ M$_\odot$ down to $3.8 \times 10^{11}$ M$_\odot$. These shifts in information relevant to structure formation do also appear to have a physical basis in the structure distributions of each of the simulations. The GR simulation contains ∼ 115 000 structures at $z = 0$, compared to ∼ 101 000 for F4 and ∼ 109 000 for N1. Increasing the strength of the fifth force in *f(R)* and nDGP models is known to enhance the abundance of massive haloes at the expense of smaller structures, with emptier voids between structures (Schmidt et al. 2008; Mitchell et al. 2021). The authors' conjecture that this behaviour is the cause of the movement of the feature importance peaks in Fig. 4, as this condensing of density signals localizes information.

Fig. 5 shows this effect on IN class membership in the GR and F4 simulations, respectively. As these initial conditions are smoothed to generate the machine learning features, larger radial (and therefore mass) scales are required to capture all the particles that end up in structures in GR. However in F4, the IN class particles are more local, due to the effects of the fifth force described above. Lower mass scales are sufficient to capture all of the particles that end up in a given structure, potentially explaining the movement of the feature importance peaks in Fig. 4 as the strength of the modified gravity theories increases.

---

[4] Higher sample sizes may be used but yield diminishing additional returns on model performance.

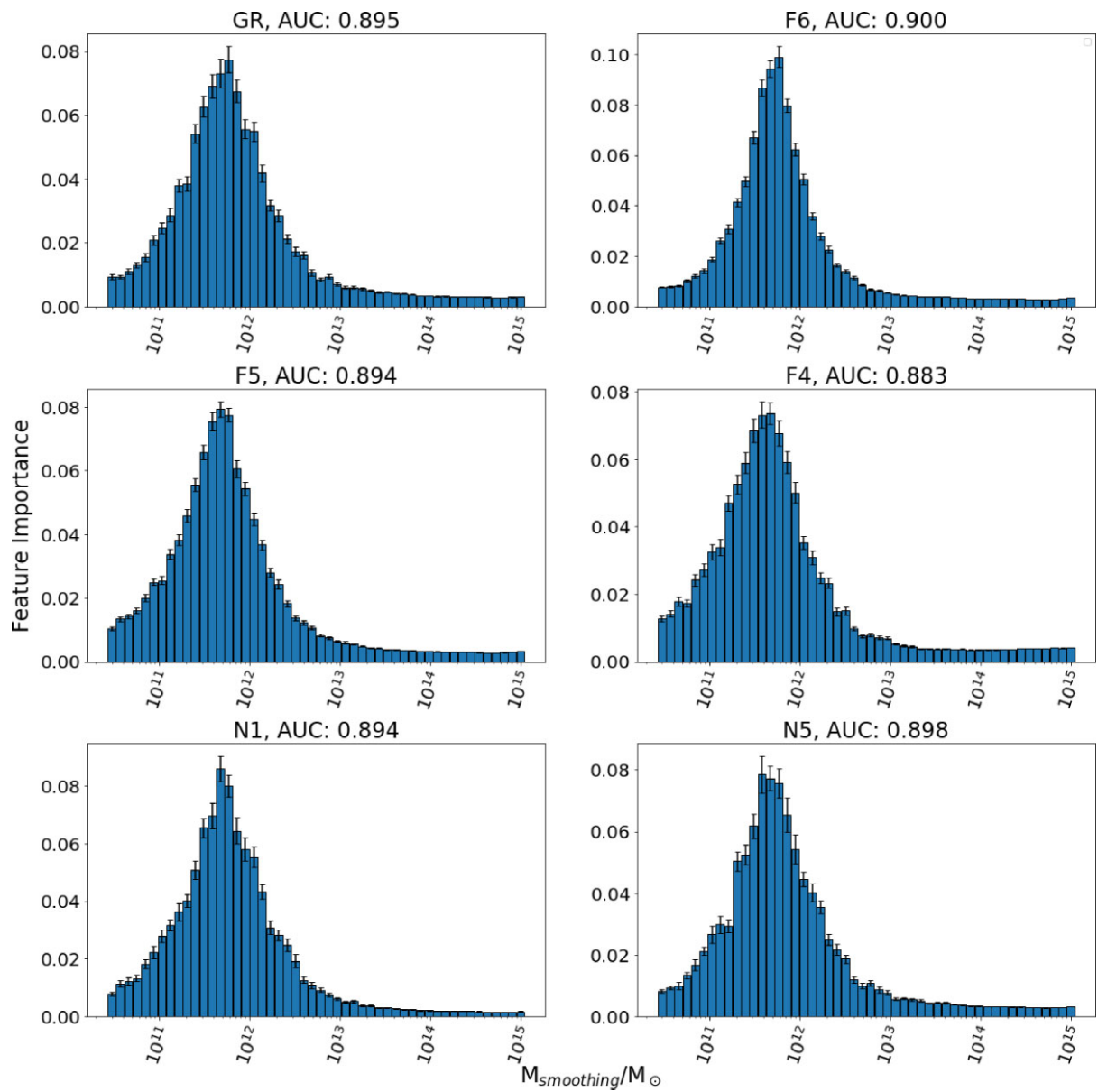

**Figure 4.** Feature importance across a sample of models with their hyperparameters optimized and trained on GR, F6, F5, F4, N1, and N5, respectively. 50 training data sets are generated for each of GR, F6, F5, F4, N1, and N5. Each of these data sets is used to train a classifier, leading to a sample of 50 models per cosmology. The average AUC score for each set of models is also given as an indicator of their performance.

As the AUC scores show, even though the difference between theories is captured in the varied mass scales at which they decide structure membership, the density contrast provides just as complete a description of structure formation for any of the models trained on their respective gravities as it does for GR. It should be noted that whilst the relative motion of the feature importance peak between cosmologies generated from the same initial conditions is indicative of mechanical changes, its position per set of simulations will change depending on resolution.

### 5.2 Generalizability of the $\Lambda$CDM model

Further to the properties of models trained on individual simulations, the GR model itself also seems to generalize well to other gravities. Fig. 6 shows how the performance of the GR model varies as it is applied to non-GR data. 50 data sets of 50 000 particles were generated for each modified gravity. Each of these data sets was passed to the GR model and compared with ground truth to build a probability-distribution-function (PDF) of the GR model's performance against each other gravity.

For $f(R)$, the spherical collapse model trained on GR data performs well across all three simulations. However, Fig. 6 does show that this model does not perform as well on data from F4. This suggests that the spherically aggregated information from the initial density field is less helpful in determining structure membership in the final conditions for F4 gravity. For F5 and F6, the performance of the GR trained ML model is consistent with its performance on GR data. Some intermediary simulation data that resides between F4 and F5 would provide the opportunity to probe this behaviour further. However, these results certainly seem to indicate that an ML model trained on Press–Schechter spherical collapse in GR generalizes well for $f(R)$ gravity for models with $|f_{R0}| < 10^{-4}$.

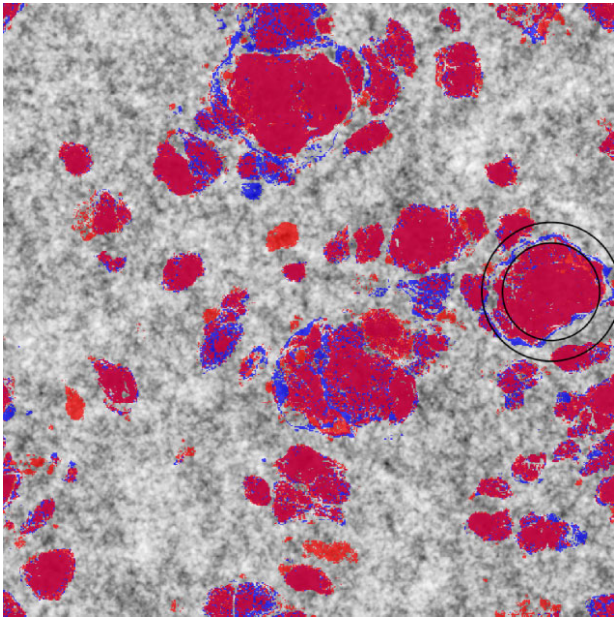

**Figure 5.** A slice of the initial density field, with IN class particles from GR in blue, and IN class particles from F4 in red. The black circles show a sample of smoothing radii from the feature engineering process.

For nDGP gravity, the spherical collapse GR model generalizes well to both N1 and N5, despite N1 violating the $H_0 r_c > 3$ constraint (Lombriser et al. 2009). Some divergence is beginning to occur as the strength of $H_0 r_c$ increases from N5 to N1, but not to the same pronounced degree as is evident in $f(R)$. Violating constraints on nDGP gravity seems to have a weaker impact on the generalizability of spherical collapse. However, there are frameworks for constraining nDGP gravity that are still awaiting the results of ongoing and upcoming sky surveys (Mitchell et al. 2021). The outcomes of these projects may provide more insight into the sensitivity of constraints in nDGP and this result, but the ML model that has learned structure formation from spherical collapse in GR certainly seems to generalize well in spite of current constraints.

## 6 CONCLUSIONS

In this work, we trained a random-forest classifier on *N*-body simulations of two modified gravity theories to find a relationship between the initial conditions of the density field at high-redshift ($z = 99$) and the present day. This approach has been found valuable in the context of simulations in $\Lambda$CDM (Lucie-Smith et al. 2018). In particular, it showed that in $\Lambda$CDM the initial density field contains sufficient information in order to predict the formation of DM haloes. The accuracy was shown to match that of (semi-)analytical frameworks based on the spherical collapse model. The main result of this paper is that these frameworks remain a very good description also in the screened modified gravity theories considered here, namely $f(R)$ theories and nDGP models.

The ML models constructed for this work also captured other modified gravity behaviours. First, they seemed to show the effects of changes in clustering behaviour between GR and the two screened gravity theories. The suppression of small clusters, enhancement of voids and larger structures, and the localizing of density signals was captured by the ML models as the strength of the fifth force increased. This was shown in the feature importance of their structure formation decisions. Secondly, the performance of the GR model maps well to established constraints on the strength of the fifth force. When the screening mechanisms were tuned beyond constraints that are known to impact clustering, the performance of the GR model decreased in a statistically significant manner in the $f(R)$ case. In the nDGP case, the performance decrease was less obvious but still present. The fact that the ML models elucidated these behaviours also shows that the key result, the generalizability of spherical collapse to screened modified gravity, is far more than a statistical triviality. Both the feature importance results and the GR model performance results form natural targets for further work, such as testing the impact of scalar field information on structure membership. The methodology described in this work may also generalize to other modified gravities with scalable extra forces, expanding structure formation as a natural test bed for new theories of gravity. Finally, we hope the results of this work lend more weight to the use of machine learning as a tool to do meaningful cosmology.

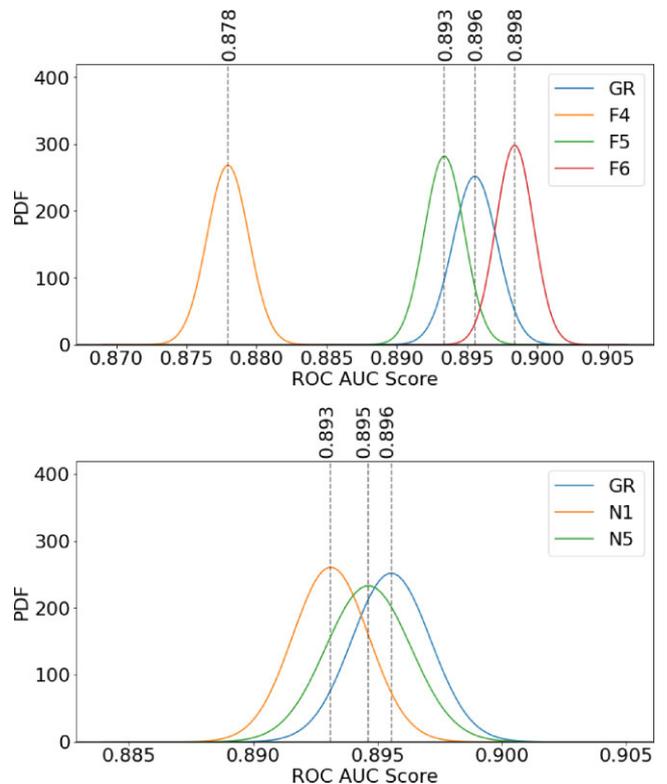

**Figure 6.** PDFs over the ROC AUC performance metric generated from applying a GR classifier to data sets generated from each of the other modified gravities. 50 test data sets were generated for each of GR, F6, F5, F4, N1, and N5. These batches of test data were then handed over to a model trained using GR data for inference. The performance of the GR model over each of these batches was used to generate a PDF for each gravity. The GR model seems to perform largely identically on all gravities except F4.

## ACKNOWLEDGEMENTS

JB thanks Luisa Lucie-Smith for useful conversations on the conversion of her pipeline for use in this work. JB would also like to give special thanks to Adam Seed and James Coyle for helpful comments and feedback on the development of the data pipeline. We are also grateful to Adam Moss for insightful comments on testing our pipeline and Hans Winther for providing testing simulations. CvdB was supported (in part) by the Lancaster–Manchester–Sheffield Consortium for Fundamental Physics under STFC grant: ST/T001038/1. CA and BL were supported by an ERC Starting Grant, ERC-StG-PUNCA-716532. BL is additionally supported by the STFC consolidated grants [ST/P000541/1, ST/T000244/1]. This work used the DiRAC@Durham facility managed by the Institute for Computational Cosmology on behalf of the STFC DiRAC HPC Facility (www.dirac.ac.uk). The equipment was funded by BEIS capital funding via STFC capital grants ST/K00042X/1, ST/P002293/1, ST/R002371/1, and ST/S002502/1, Durham University and STFC operations grant ST/R000832/1. DiRAC is part of the National e-Infrastructure.

## DATA AVAILABILITY

The data underlying this article were generated by CA and BL at the DiRAC@Durham facility, with relevant details above. The derived data generated in this research will be shared on reasonable request to the corresponding author.

## APPENDIX A: THE IMPACT OF MODIFIED GRAVITY ON GRAVITATIONAL BINDING

Pertaining to the gravitational binding condition used in our structure definition in Section 4.1, it is reasonable to question the appropriateness of this assumption in the context of modified gravities, particularly $f(R)$ gravity. A more appropriate model for the modified gravitational potential could be $U = (1 + \beta^2)U_G$, where $U_G$ is the standard Newtonian potential and $\beta = \sqrt{1/6}$, which takes into account the impact of the fifth force appearing in $f(R)$ theories. When defining haloes in the final conditions, all particles are bound based on the friends-of-friends method of the associated halo finder. There is then an unbinding process, where the kinetic energy of the particle is compared to the gravitational potential of the relevant structure. When deciding gravitational unbinding, using $U$ rather than $U_G$ could lead to more particles being included in structures in $f(R)$. This effect would be most prevalent in F4 gravity, where all structures are unscreened. In order to make sure that the results of Sections 5.1 and 5.2 are not impacted by the fifth force in gravitational binding, we repeated those tests with the original F4 data and a new F4 halo catalogue generated using *U* for binding, denoted here as F4M. Figs A1 and A2 below show the feature importance of F4 and F4M models, and the GR model performance test results analogous to those in Section 5.2.

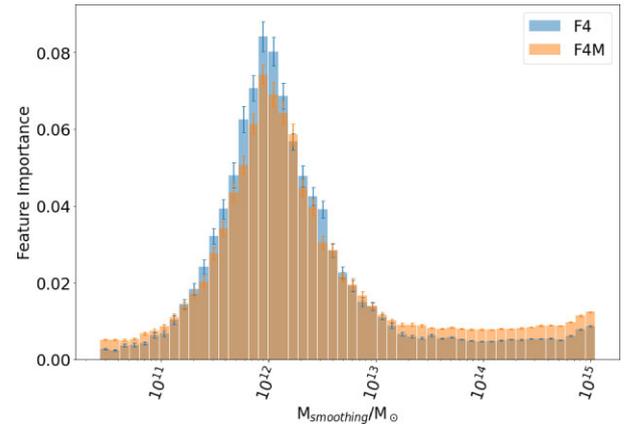

**Figure A1.** Feature importance for models trained on F4 (blue) and F4M (orange) data.

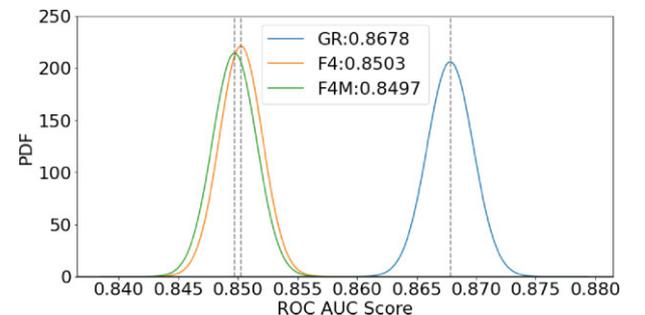

**Figure A2.** PDFs over the ROC AUC performance metric generated from applying a GR classifier to data sets generated from GR, F4, and F4M.

Fig. A1 shows that the F4M feature importance is spread more broadly over the whole mass range, but the peak is in the same location, with the same behaviour as F4. Fig. A2 shows that the GR model performs almost identically (a 0.06 per cent difference) on F4M data as on F4 data. These results imply that using the GR potential in the halo-finder for gravitational unbinding does not impact the veracity of the results of this work, and the behaviours observed by this ML study do not arise from this definition of gravitational binding.

This paper has been typeset from a TEX/LATEX file prepared by the author.